\documentclass[aps,prb,twocolumn,groupedaddress,showpacs]{revtex4-1}

\usepackage{graphicx}
\usepackage{amsfonts}
\usepackage{amsmath}
\usepackage{amssymb}
\usepackage{bm}

\begin{document}
\title{Peculiar long-range supercurrent in SFS junction containing a noncollinear magnetic domain in the ferromagnetic region}
\author{Hao Meng}
\email{menghao1982@shu.edu.cn}
\author{Xiuqiang Wu}

\affiliation{National Laboratory of Solid State Microstructures and Department of Physics, Nanjing University, Nanjing 210093, China}
\date{\today }

  \begin{abstract}
   We study the supercurrent in a superconductor-ferromagnet-superconductor heterostructure containing a noncollinear magnetic domain in the ferromagnetic region. It is demonstrated that the magnetic domain can lead to a spin-flip process, which can reverse the spin orientations of the singlet Cooper pair propagating through the magnetic domain region. If the ferromagnetic layers on both sides of magnetic domain have the same features, the long-range proximity effect will take place. That is because the singlet Cooper pair will create an exact phase-cancellation effect and gets an additional $\pi$ phase shift as it passes through the entire ferromagnetic region. Then the equal spin triplet pair only exists in the magnetic domain region and can not diffuse into the other two ferromagnetic layers. So the supercurrent mostly arises from the singlet Cooper pairs and the equal spin triplet pairs are not involved. This behavior is quite distinct from the common knowledge that long-range supercurrent induced by inhomogeneous ferromagnetism stems from the equal spin triplet pairs. The result we presented here provides a new way for generating the long-range supercurrent.
  \end{abstract}

 \pacs{74.78.Fk, 73.40.-c, 74.50.+r, 73.63.-b} \maketitle

  The interplay between superconductivity and ferromagnetism in mesoscopic structures has been extensively studied because of the underlying rich physics and potential applications in spintronics and quantum information~\cite{Buz,BerRMP,AAGolubov,Esc}. When a homogeneous ferromagnet ($F$) is sandwiched between two s-wave superconductors ($S$) to form a Josephson junction, the magnetic configuration of the $F$ layer may substantially modify the spatial properties of the superconducting order parameter. This behavior is induced by the different action of the ferromagnetic exchange field $h$ on the spin-up and spin-down electrons that form the Cooper pair. Then this pair inside the $F$ layer acquire a relative phase shift $Q\cdot{R}$, where $Q\simeq2h/\hbar{v_F}$, $v_F$ is the Fermi velocity and $R$ is the thickness of the $F$ layer. This phase shift changes with $R$ and results in an oscillation of critical current accompanied with a rapid decay. The above oscillation will lead to the transition from the so-called $0$ state to the $\pi$ state~\cite{Buz,BerRMP,AAGolubov,Esc}.

  Two kinds of approaches have been proposed to produce long-range supercurrent in $SFS$ junction. The first approach requires inhomogeneous magnetism in $F$ region so that the interested equal spin triplet pairs can be generated~\cite{BerRMP}. One way to achieve this purpose is by arranging ferromagnetic trilayer with noncollinear magnetizations~\cite{Hou}. In this geometry, the spin-flip processes at the interface $F$ layers can convert the singlet Cooper pairs into the equal spin triplet pairs~\cite{Esc,Eschrig}. The triplet pairs can penetrate into the central $F$ layer over a long distance unsuppressed by the exchange interaction so that the proximity effect is enhanced. The second approach requires the $F$ layer to be arranged antiparallel. This situation was described by Blanter \emph{et al.} for a $SFFS$ junction~\cite{YMBlanter}. The physical origin of the enhanced proximity effect is described as a compensation of the relative phase shift of a Cooper pair as it passes from the first $F$ layer into the second one. If the two $F$ layers have the same thickness, the net change in the relative phase of the Cooper pair is zero in the clean limit. This enhanced Josephson current has been proved by recent experiment~\cite{JWARobinson}.

\begin{figure}
  \centering
  \includegraphics[width=3.1in]{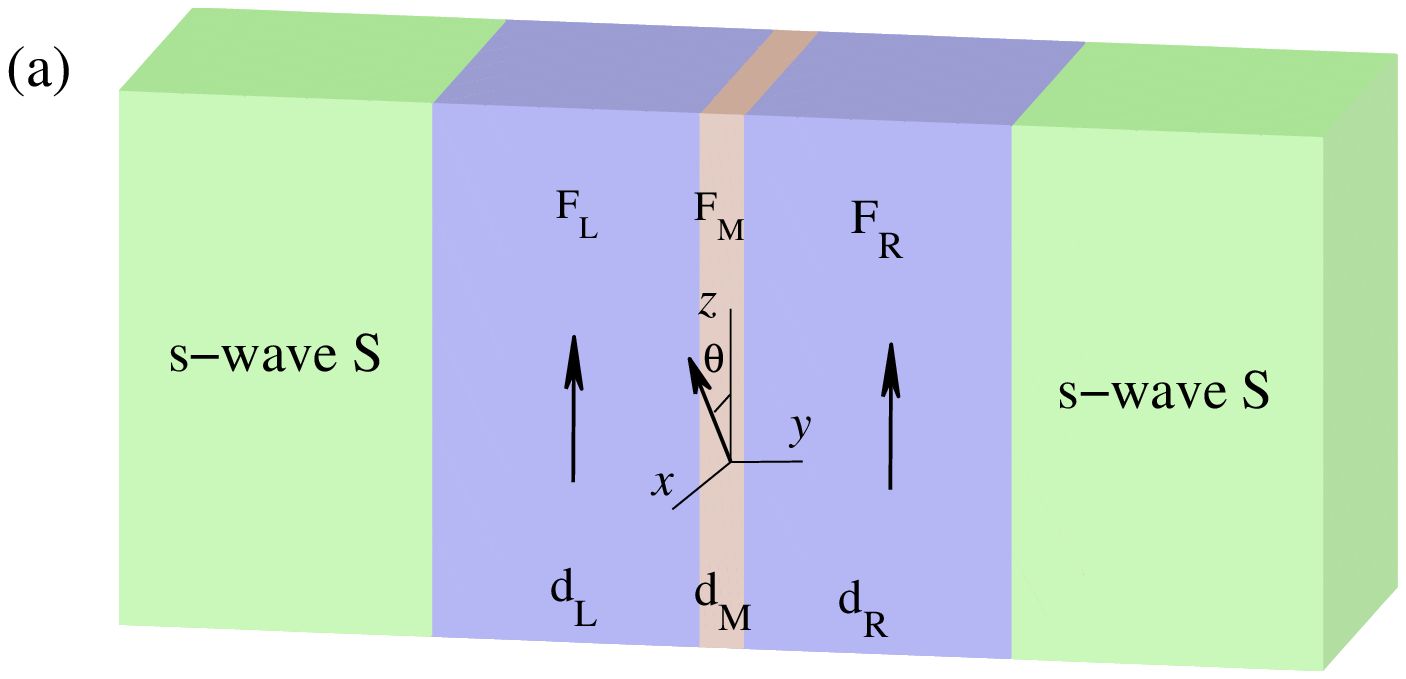} 
  \includegraphics[width=3.3in]{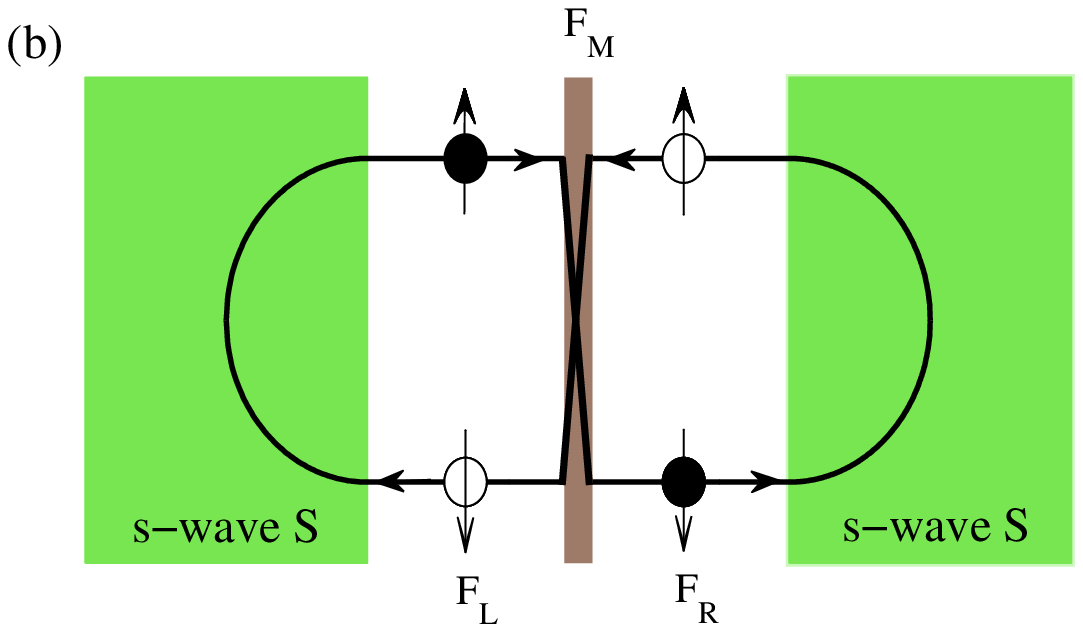} 
  \caption{(color online) (a) Schematic diagram of $SFS$ structure with two ferromagnetic layers $F_L$ and $F_R$ oriented along the $z$ axis and separated by a noncollinear magnetic domain $F_M$. The lengths of $F_L$, $F_R$ and $F_M$ are denoted by $d_L$, $d_R$ and $d_M$, respectively. The phase difference between the two $S$s is $\phi=\phi_R-\phi_L$. (b) The transmission of electron and hole in above structure.}
  \label{fig.1}
  \end{figure}

  In this paper, we predict the third approach to generate the long-range supercurrent in $SFS$ structure, shown schematically in figure~\ref{fig.1}(a). The ferromagnetic region consists of two ferromagnetic layers ($F_L$ and $F_R$) with magnetizations oriented in same directions. The $F_L$ layer and $F_R$ layer are separated by a clean magnetic domain ($F_M$), whose magnetization is misaligned with direction of the $F_L$ layer and $F_R$ layer. The magnetic domain $F_M$ can induce a spin-flip process, which reverses the spin orientations of the singlet Cooper pair propagating through the $F_M$ region. This process will make the singlet Cooper pair create a phase-cancellation effect and obtain an additional $\pi$ phase shift. If the $F_L$ layer and $F_R$ layer have the same thickness and exchange field, the net change in the relative phase of a singlet Cooper pair is $\pi$ when it passes through the entire $F$ region. In this case, the contribution to the long-range supercurrent mostly arises from the singlet Cooper pairs. This is because the equal spin triplet pairs only display in $F_M$ region and can not diffuse into the $F_L$ and $F_R$ layers.

  In our numerical calculation, the transport direction is along the \emph{y} axis, and the system is assumed to be infinite in the \emph{x-z} plane. The BCS mean-field effective Hamiltonian~\cite{Buz,Gen} is
  \begin{equation}
  \label{Eq1}
  \begin{aligned}
    H_{eff}&=\int{d\vec{r}}\{\sum_{\alpha,\beta}\psi^{\dag}_{\alpha}(\vec{r})[H_e(\hat{\textbf{1}})_{\alpha\beta}-(\vec{h}\cdot\vec{\sigma})_{\alpha\beta}]\psi_{\beta}(\vec{r})\\
   &+\frac{1}{2}[\sum_{\alpha,\beta}(i\sigma_{y})_{\alpha\beta}\Delta(\vec{r})\psi^{\dag}_{\alpha}(\vec{r})\psi^{\dag}_{\beta}(\vec{r})+h.c.]\},
  \end{aligned}
  \end{equation}
  where $H_e=-\hbar^2\nabla^{2}/2m-E_F$, $\psi^{\dag}_{\alpha}(\vec{r})$ and $\psi_{\alpha}(\vec{r})$ are creation and annihilation operators with spin $\alpha$. $\hat{\sigma}$ and $E_F$ are Pauli matrices and the Fermi energy, respectively. The superconducting gap is given by $\Delta(\vec{r})=\Delta(T)[e^{i\phi_{L}}\Theta(-y)+e^{i\phi_{R}}\Theta(y-d_{F})]$ with $d_{F}=d_{L}+d_{M}+d_{R}$. Here, $\Delta(T)$ accounts for the temperature-dependent energy gap. It satisfies the BCS relation $\Delta(T)=\Delta_0\tanh(1.74\sqrt{T_c/T-1})$ with $T_c$ the superconducting critical temperature. $\Theta(y)$ is the unit step function, and $\phi_{L(R)}$ is the phase of the left (right) $S$. The exchange field $\vec{h}$ due to the ferromagnetic magnetizations in the $F$ region is described by
  \begin{equation}
  \label{Eq2}
  \vec{h}=
  \left\{
  \begin{array}{lcl}
   h_{L}\hat{z}, && 0<y<d_{L} \\
   h_{M}(\sin\theta\hat{x}+\cos\theta\hat{z}), && d_{L}<y<d_{L}+d_{M} \\
   h_{R}\hat{z}, && d_{L}+d_{M}<y<d_{F},
   \end{array}
   \right.
  \end{equation}
  where $\theta$ is the misorientation angle of magnetization in the magnetic domain $F_M$ region. Based on the Bogoliubov transformation $\psi_{\alpha}(\vec{r})=\sum_{n}[u_{n\alpha}(\vec{r})\hat{\gamma}_{n}+v^{\ast}_{n\alpha}(\vec{r})\hat{\gamma}^{\dag}_{n}]$ and the anticommutation relations of the quasiparticle annihilation and creation operators $\hat{\gamma}_{n}$ and $\hat{\gamma}^{\dag}_{n}$, we have the Bogoliubov-de Gennes (BdG) equation~\cite{Buz,Gen}
  \begin{equation}
  \label{Eq3}
  \left(
  \begin{array}{ccc}
  \hat{H}(y) & \hat{\Delta}(y) \\
	-\hat{\Delta}^{*}(y) & -\hat{H}^{*}(y)\\
  \end{array}
  \right)
  \left(
  \begin{array}{ccc}
    \hat{u}(y) \\
	\hat{v}(y) \\
  \end{array}
   \right)
   =E
   \left(
  \begin{array}{ccc}
   \hat{u}(y)  \\
	\hat{v}(y) \\
  \end{array}
  \right),
  \end{equation}
  where $\hat{H}(y)=H_{e}\hat{\textbf{1}}-h_{z}(y)\hat{\sigma}_{z}-h_{x}(y)\hat{\sigma}_{x}$ and $\hat{\Delta}(y)= i\hat{\sigma}_{y}\Delta(y)$. Here $\hat{u}(y)=(u_{\uparrow}(y)$, $u_{\downarrow}(y))^{T}$ and $\hat{v}(y)=(v_{\uparrow}(y)$, $v_{\downarrow}(y))^{T}$ are two-component wave functions.

  Since the transversal momentum components are conserved, we consider the configuration in the one dimensional regime for simplicity. The BdG equation can be easily solved for each superconducting electrode and each $F$ region, respectively. The scattering problem can be solved by considering the boundary conditions at the interfaces. Each interface gives a scattering matrix. The total scattering matrix of the system can be obtained by the combination of all these scattering matrices of the interfaces. From the total scattering matrix, we can obtain the Andreev reflection amplitudes $a_1\sim{a_4}$ of the junction where $a_{1(2)}$ is for the reflection from an electron-like to a hole-like quasiparticle with spin up (down), and $a_{3(4)}$ is for the reverse process with spin up (down)~\cite{Meng,JunFeng}. The dc Josephson current can be expressed in terms of the Andreev reflection amplitudes by using the finite-temperature Green's function formalism~\cite{Blo,Fur,Zhe,Tan}
  \begin{equation}
  \label{Eq4}
  \begin{aligned}
  &I_{e}(\phi)=\frac{k_BTe\Delta}{4\hbar}\sum_{k_{\parallel}}\sum_{\omega_{n}}\frac{k_{e}(\omega_{n})+k_{h}(\omega_{n})}{\Omega_{n}}\cdot\\
  &[\frac{a_1(\omega_{n},\phi)-a_2(\omega_{n},\phi)}{k_{e}}+\frac{a_3(\omega_{n},\phi)-a_4(\omega_{n},\phi)}{k_{h}}],
  \end{aligned}
  \end{equation}
  where $\omega_{n}=\pi{k_B}T(2n+1)$ are the Matsubara frequencies with $n=0, 1, 2,\ldots$ and $\Omega_{n}=\sqrt{\omega^{2}_{n}+\Delta^{2}(T)}$. $k_e(\omega_{n})$, $k_h(\omega_{n})$, and $a_j(\omega_{n},\phi)$ with $j=1, 2, 3, 4$ are obtained  from $k_e$, $k_h$, and $a_j$ by analytic continuation $E\rightarrow{i}\omega_{n}$. $k_{e(h)}$ is the wave vector for electron (hole) in the $S$s. Then the critical current is derived from $I_c=max_{\phi}|I_e(\phi)|$.

  In principle, to obtain the spin singlet pair and the spin triplet pair amplitude functions as well as the local density of the states (LDOS), we need to solve the BdG equation~(\ref{Eq3}) by Bogoliubov's self-consistent field method~\cite{Gen,Ketterson,Halterman}. We put the junction in a one dimensional square potential well with infinitely high walls. Accordingly, the solution in equation~(\ref{Eq3}) can be expanded in terms of a set of basis vectors of the stationary states~\cite{Landau}, $u_{\alpha}(\vec{r})$$=$$\sum_{q}u^{\alpha}_{q}\zeta_{q}(y)$ and $v_{\alpha}(\vec{r})=\sum_{q}v^{\alpha}_{q}\zeta_{q}(y)$ with $\zeta_{q}(y)=\sqrt{2/d}\sin(q{\pi}y/d)$. Here $q$ is a positive integer and $d=d_{S1}+d_{F}+d_{S2}$. $d_{S1}$ and $d_{S2}$ are the length of the left and right $S$s, respectively. The BdG equation~(\ref{Eq3}) is solved iteratively together with the self-consistency condition $\Delta(y)=g(y)f_3(y)$~\cite{Gen}. Here the singlet pair amplitude is give by~\cite{Halterman}
   \begin{equation}
   \label{Eq5}
   f_3(y)=\frac{1}{2}\sum_{0\leq{E}\leq\omega_{D}}[u_{\uparrow}(y)v^{*}_{\downarrow}(y)-u_{\downarrow}(y)v^{*}_{\uparrow}(y)]\tanh(\frac{E}{2k_BT}),
   \end{equation}
  where $\omega_{D}$ is the Debye cutoff energy. The effective attractive coupling $g(y)$ will be taken to be zero outside the $S$ and a constant within it. Iterations are performed until self-consistency is reached, starting from the stepwise approximation for the pair potential. The spin triplet pair with zero spin projection and the equal spin triplet pair amplitude functions are defined as follows~\cite{Halterman}
  \begin{equation}
  \label{Eq6}
  f_{0}(y,t)=\frac{1}{2}\sum_{E>0}[u_{\uparrow}(y)v^{*}_{\downarrow}(y)+u_{\downarrow}(y)v^{*}_{\uparrow}(y)]\eta(t),
  \end{equation}
  \begin{equation}
  \label{Eq7}
  f_{1}(y,t)=\frac{1}{2}\sum_{E>0}[u_{\uparrow}(y)v^{*}_{\uparrow}(y)-u_{\downarrow}(y)v^{*}_{\downarrow}(y)]\eta(t),
  \end{equation}
  where $\eta(t)=\cos(Et)-i\sin(Et)\tanh(E/2k_BT)$. The above singlet and triplet pair amplitudes are all normalized to the value of the singlet pairing amplitude in a bulk $S$ material. The LDOS is given by~\cite{Halterman}
  \begin{equation}
  \label{Eq8}
  N(y,\epsilon)=-\sum_{\alpha}[u^{2}_{\alpha}(y)f'(\epsilon-E)+v^{2}_{\alpha}(y)f'(\epsilon+E)],
  \end{equation}
  where $f'(\epsilon)=\partial{f}/\partial{\epsilon}$ is the derivative of the Fermi function. The LDOS is normalized by its value at $\epsilon=3\Delta_{0}$ beyond which the LDOS is almost constant.

  Unless otherwise stated, we use the superconducting gap $\Delta_0$ as the unit of energy. The Fermi energy is $E_F=1000\Delta_0$ and temperature is $T/T_c=0.1$. In self-consistent field method, we consider the low temperature limit and take $k_Fd_{S1}$=$k_Fd_{S2}$=400, $\omega_{D}/E_F$=0.1, the other parameters are the same as the ones mentioned above.

  \begin{figure}
  \centering
  \includegraphics[width=3.3in]{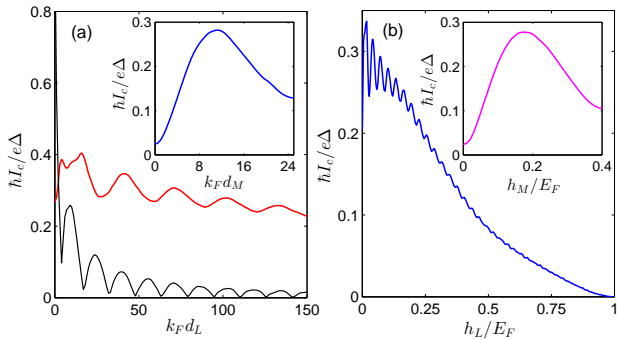} 
  \caption{(Color online) (a) Critical current as a function of length $k_Fd_{L} (=k_Fd_{R})$ for exchange field $h_M/E_F=0$ (black line) and $h_M/E_F=0.17$ (magenta line) then $k_Fd_{M}=10$, and inset shows the critical current versus $k_Fd_{M}$ for $k_Fd_{L}=k_Fd_{R}=100$. Parameters used in (a): $h_L/E_F=h_R/E_F=0.1$ and $h_M/E_F=0.17$. (b) Critical current as a function of exchange field $h_L/E_F (=h_R/E_F)$ for $h_M/E_F=0.17$, and inset shows the critical current versus $h_M/E_F$ for $h_L/E_F=h_R/E_F=0.1$. Parameters used in (b): $k_Fd_{L}=k_Fd_{R}=100$ and $k_Fd_{M}=10$. In all plots $\theta=\pi/2$.}
  \label{fig.2}
  \end{figure}

   In figure~\ref{fig.2}(a), we present the dependence of the critical current $I_c$ on the length $k_Fd_{L} (=k_Fd_{R})$ for two different exchange fields of $F_M$ region. It is well known that, if $h_M/E_F=0$ the spin-flip process does not occur in $F_M$ region, the critical current $I_c$ exhibits oscillations with a period $2\pi\xi_{F}$ and simultaneously decays exponentially on the length scale of $\xi_{F}$~\cite{Buz}. Here, $\xi_{F}$ is the magnetic coherence length. The main reason is described below: a Cooper pair entering into the $F$ layer receives a finite momentum $Q$ from the spin splitting of up and down bands. So the spin singlet pairing state $\mid\uparrow\downarrow\rangle-\mid\downarrow\uparrow\rangle$ in $S$ will be converted into the mixed state $\mid\uparrow\downarrow\rangle{e^{iQ\cdot{R}}}-\mid\downarrow\uparrow\rangle{e^{-iQ\cdot{R}}}$ in $F$ layer, thus leading to a modulation of the pair amplitude with the thickness $R$ of $F$ layer. Then the $0-\pi$ transition will arise due to spatial oscillations of the pair amplitude. In this case the phase shift induced by the $F$ layer is additive and the relative phase is generally nonzero so that the supercurrents are suppressed.

   In contrast, when $h_M/E_F=0.17$, $I_c$ will slowly decrease with the thickness $k_Fd_{L}$. The electron and hole transport process is shown in figure~\ref{fig.1}(b). Because the magnetization direction of the $F_M$ region is along the $\emph{x}$ axis ($\theta=\pi/2$), which is orthogonal to the magnetization in $F_L$ layer and $F_R$ layer, the spin-flip process will appear in the $F_M$ region. As a result, when a electron $\mid\uparrow\rangle_e$ transmits from $F_L$ layer to $F_R$ layer, the spin-flip can convert $\mid\uparrow\rangle_e$ into $\mid\downarrow\rangle_e$. Subsequently, the $\mid\downarrow\rangle_e$ is Andreev reflected at $F_RS$ interface and will be further converted into hole $\mid\uparrow\rangle_h$. The $\mid\uparrow\rangle_h$ moving to left is consequently inverted to $\mid\downarrow\rangle_h$. Finally, this $\mid\downarrow\rangle_h$ will propagate to the $SF_L$ interface and be reflected back as the original $\mid\uparrow\rangle_e$. Hence, the spin-flip process occurring in the $F_M$ region can reverse the spin orientations of the electron and Andreev-reflected hole transporting between $F_L$ layer and $F_R$ layer. While if the mixed state $\mid\uparrow\downarrow\rangle{e^{iQ\cdot{R}}}-\mid\downarrow\uparrow\rangle{e^{-iQ\cdot{R}}}$ passes from $F_L$ layer into $F_R$ layer, it will be converted into a new state:
   \begin{equation}
   \label{Eq9}
   \begin{aligned}
   &\mid\downarrow\uparrow\rangle{e^{iQ'\cdot{R'}}}-\mid\uparrow\downarrow\rangle{e^{-iQ'\cdot{R'}}}\\
   &=-(\mid\uparrow\downarrow\rangle{e^{-iQ'\cdot{R'}}}-\mid\downarrow\uparrow\rangle{e^{iQ'\cdot{R'}}})\\
   &=\mid\uparrow\downarrow\rangle{e^{i(-Q'\cdot{R'}+\pi)}}-\mid\downarrow\uparrow\rangle{e^{-i(-Q'\cdot{R'}+\pi)}}.
   \end{aligned}
   \end{equation}
   Therefore, when the Cooper pair passes through the $F_L$ layer, it will acquire a relative phase shift $\delta\chi_1=Qd_{L}$. Similarly, traversing through the $F_R$ layer, it gets the other phase shift $\delta\chi_2=-Q'd_{R}+\pi$. As a result, this situation can be described as a superposition of the phase shift of the Cooper pair as it travels through the entire $F$ region: $\chi=\delta\chi_1+\delta\chi_2$. If the $F_L$ and $F_R$ layer have the same exchange field and thickness, the net change in the relative phase of the Cooper pair is $\chi=\pi$, and it will not turn to $0$ with increase of the length of $F_L$ layer and $F_R$ layer. Provided one does not take absolute value for $I_e(\phi)$ to define the critical current $I_c$, $I_c$ is always negative and is correspond to the $\pi$ state. As the magenta line shown in figure~\ref{fig.2}(a), $I_c$ also displays an oscillatory behavior with increasing the length $k_Fd_{L}$. From this behavior, it is easier to deduce that the spin orientations of the part of Cooper pairs will be inverted by spin-flip in $F_M$ region, so that these pairs can lead to a substantially enhanced critical current but do not provide the oscillations for this current. However, the rest of Cooper pairs pass through the $F_M$ region without spin-flip and can not generate the cancellation of the relative phase. Consequently, the transmission of these pairs makes the critical current oscillate with the length of entire $F$ region. In addition, another interesting property is the nonmonotonic dependence of the critical current $I_c$ as a function of the length $k_Fd_{M}$ (see the inset in figure~\ref{fig.2}(a)). We find that the maximum of $I_c$ is nearly located at $k_Fd_{M}=10$. It means that the spin-flip ratio reaches their maximal value in this condition.

   \begin{figure}
   \centering
   \includegraphics[width=3.3in]{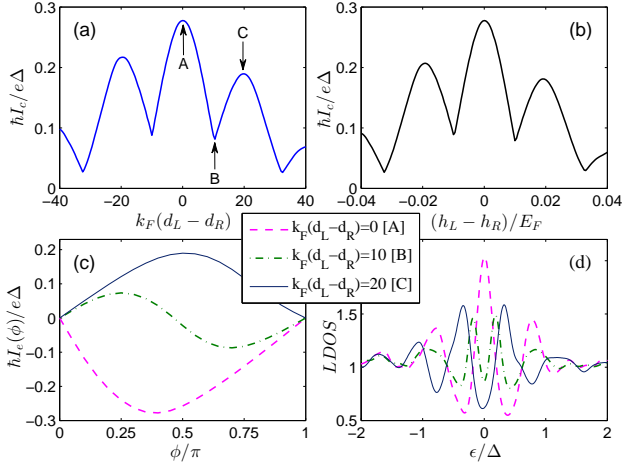} 
   \caption{(Color online) Critical current (a) as a function of the length difference $k_F(d_{L}-d_{R})$ for $h_L/E_F=0.1$, and (b) as a function of the exchange field difference $(h_L-h_R)/E_F$ for $k_Fd_{L}=100$. The current-phase relation $I_e(\phi)$ (c) and the averaged LDOS in all $F$ region (d) corresponding to the point A, B and C in panel (a). The parameters in $F_M$ and $F_R$ layers have the fixed values $k_Fd_{M}=10$, $k_Fd_{R}$=100, $h_M/E_F=0.17$, $h_R/E_F=0.1$ and $\theta=\pi/2$. Here the LDOS is calculated at $k_BT=0.001$.}
   \label{fig.3}
   \end{figure}

   Next, we discuss the dependence of critical current $I_c$ on the exchange field $h_L/E_F(=h_R/E_F)$. As plotted in figure~\ref{fig.2}(b), $I_c$ almost decreases monotonically to 0 with the increase of $h_L/E_F$. This is easily understood, because normal Andreev reflection occurred at the $SF_L$ and $F_RS$ interface will be suppressed by exchange splitting of $F_L$ layer and $F_R$ layer. Especially, if $h_L/E_F=h_R/E_F=1$, the $F_L$ layer and $F_R$ layer are all converted into half metal and only one spin band can be occupied, then the Andreev reflections at the interfaces will be completely prohibited. In this case, none of Cooper pairs can transmit from the left $S$ to the right one, so the Josephson current would be complete suppressed. This feature further demonstrates that the current mostly arises from the contribution of the singlet Cooper pairs but not the equal spin triplet pairs. That is because the equal spin triplet pairs can penetrate over a long distances into the half metal and will scarcely be affected by exchange splitting~\cite{Eschrig,Keizer}. In addition, the inset in figure~\ref{fig.2}(b) shows the $h_M/E_F$ dependence of the critical current $I_c$. It also displays a nonmonotonic behavior as the $h_M/E_F$ is increased, and the maximum is nearly seated at $h_M/E_F=0.17$.

   To further demonstrate the conclusion mentioned beforehand, we now discuss intriguing influence of the length and exchange field on the critical current $I_c$ when the $F_L$ layer and $F_R$ layer have nonidentical physical features. As illustrated in figure~\ref{fig.3}(a) and (b), we present the dependence of $I_c$ on $k_F(d_{L}-d_{R})$ and $(h_L-h_R)/E_F$ respectively on condition that the exchange field and length of $F_R$ layer are all fixed. Take first one for example, we find that the dependence of $I_c$ on length difference $k_F(d_{L}-d_{R})$ look like a ``Fraunhofer pattern''. With the length difference close to 0, $I_c$ will increase and also accompanies the transition between the 0 and $\pi$ states. As mentioned above, if $F_L$ layer and $F_R$ layer have the identical exchange fields ($h_L/E_F=h_R/E_F=0.1$) but different lengths, the Cooper pair passing through the $F_L$ layer and $F_R$ layer could acquire the phase shift $\chi=Q(d_{L}-d_{R})+\pi$. So the variation of length difference can lead to the oscillation of $I_c$. On the other hand, if $F_L$ layer and $F_R$ layer have the same length ($k_Fd_{L}=k_Fd_{R}=100$) but different exchange fields, the Cooper pair can get the phase shift $\chi=(Q-Q')d_{L}+\pi$. $I_c$ will also oscillate with $(h_L-h_R)/E_F$ because of $Q\propto{h}$ (see figure~\ref{fig.3}(b)).

   In addition, the current-phase relations $I_e(\phi)$ in particular points are illustrated in figure~\ref{fig.3}(c), and corresponding LDOSs are plotted in figure~\ref{fig.3}(d). If one take the identical parameter in $F_L$ and $F_R$ layer, such as $k_FL_{L}=k_FL_{R}=100$ and $h_L/E_F=h_R/E_F=0.1$, the Cooper pairs will obtain a net phase shift $\chi=\pi$. Then we could observe a negative Josephson current and a shark zero energy conductance peak in LDOS, which indicate the junction is located in $\pi$ state. When the length difference $k_F(d_{L}-d_{R})=10$, it corresponds to the transition point between the 0 and $\pi$ states of the junction. At this critical point, the harmonic $I_1\sin\phi$ of the current ($I(\phi)=I_1\sin\phi+I_2\sin(2\phi)+...$) vanishes, and the $I_2\sin(2\phi)$ will be fully revealed. Subsequently, the sign of $I_c$ is changed with the increase of the length difference. For $k_F(d_{L}-d_{R})=20$, the junction is in the 0 state, and the LDOS at $\epsilon=0$ will be converted from the peak to a valley.

\begin{figure}
   \centering
   \includegraphics[width=3.3in]{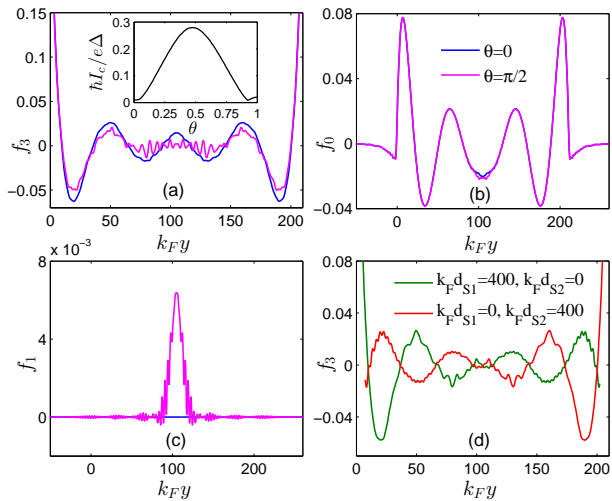} 
   \caption{(color online) Spatial distributions of the spin singlet pair amplitude $f_3$ (a), the real parts of spin triplet pair amplitude $f_0$ (b) and $f_1$ (c) for two angles $\theta=0$ and $\theta=\pi/2$ at $\omega_Dt=12$. The panels (a), (b) and (c) utilize the same legend. Inset in (a) shows the critical current as a function of the angle $\theta$. (d) The $f_3$ plotted as a function of $k_Fy$ for two cases $k_Fd_{S1}=400$, $k_Fd_{S2}=0$ and $k_Fd_{S1}=0$, $k_Fd_{S2}=400$ when $\theta=\pi/2$. Parameters used in all figures: $k_Fd_{L}=k_Fd_{R}=100$, $k_Fd_{M}=10$, $h_L/E_F=h_R/E_F=0.1$, $h_M/E_F=0.17$ and $\phi=0$. }
   \label{fig.4}
   \end{figure}

   Finally, we discuss the dependence of $I_c$ on the misorientation angle $\theta$, and the spatial distributions of the spin singlet pair and the spin triplet pair for two different $\theta$. One can see from the inset in figure~\ref{fig.4}(a) that, when the orientation of the magnetization in the $F_M$ region is perpendicular to the direction of $F_L$ layer and $F_R$ layer, the critical current reaches the maximum. However, it decreases to minimum on condition that the magnetization of $F_M$ region is parallel or antiparallel to the one in $F_L$ layer and $F_R$ layer. For given thickness of the $F_M$ region, it is possible to find the exchange field at which switching between parallel and perpendicular orientations will lead to switching of $I_c$ from near-zero to a finite value. This effect may be used for engineering cryoelectronic devices manipulating spin-polarized electrons. Furthermore, we find the spin singlet pair amplitude $f_3$ oscillates in all $F$ region at $\theta=0$. But it will be coherent counteracted at $\theta=\pi$ (see the main plot in figure~\ref{fig.4}(a)). In contrast, the spin triplet pair amplitude $f_0$ are almost identical in above two cases. For $\theta=0$, the equal spin triplet pair amplitude $f_1$ is zero in entire $F$ region. However, $f_1$ only survives in $F_M$ region and can not exist in $F_L$ layer and $F_R$ layer under the condition of $\theta=\pi$. From these consequences, we can derived that the long-range supercurrent mostly arises from the coherent propagation of $f_3$, and the contributions of $f_0$ and $f_1$ can be ruled out. To further uncover this contribution, we take into account $f_3$ induced by only one $S$. According to above theory, the spin mixed state state in $F_L$ layer is expressed as $\mid\uparrow\downarrow\rangle{e^{iQ\cdot{R}}}-\mid\downarrow\uparrow\rangle{e^{-iQ\cdot{R}}}=(\mid\uparrow\downarrow\rangle-\mid\downarrow\uparrow\rangle)\cos(Q\cdot{R})+i(\mid\uparrow\downarrow\rangle+\mid\downarrow\uparrow\rangle)\sin(Q\cdot{R})$.
   This state in $F_R$ layer can be converted as $\mid\uparrow\downarrow\rangle{e^{i(-Q'\cdot{R'}+\pi)}}-\mid\downarrow\uparrow\rangle{e^{-i(-Q'\cdot{R'}+\pi)}}=(\mid\uparrow\downarrow\rangle-\mid\downarrow\uparrow\rangle)\cos(Q'\cdot{R'}+\pi)+i(\mid\uparrow\downarrow\rangle+\mid\downarrow\uparrow\rangle)\sin(Q'\cdot{R'})$. We can see the spin singlet pair has a phase shift $\pi$ in the expression of the $F_L$ layer and $F_R$ layer. But the spin triplet pair with zero spin projection shows the same description in these two layers. These inferences are in agreement with our numerical results. As demonstrated in figure~\ref{fig.4}(d), $f_3$ shows a antisymmetric configuration around the middle $F_M$ region. As a result, the superposition of two $f_3$ stemming from the left and right $S$ will make their amplitudes cancel each other.

   In conclusion, we have studied numerically the long-range supercurrent in a $SFS$ structure including a noncollinear magnetic domain in the ferromagnetic region. We find the magnetic domain could induce a spin-flip process, which can reverse the spin orientations of the singlet Cooper pair when this pair propagate through the magnetic domain region. This process will make the singlet Cooper pair generate a phase-cancellation effect and acquire an additional $\pi$ phase shift. If the ferromagnetic layers on both sides of magnetic domain have the same features (such as thickness and exchange field), the net change in the relative phase of the singlet Cooper pair is $\pi$ when the pair passes through the entire $F$ region. In this case, the long-range supercurrent mostly stems from the singlet Cooper pairs. The reason is that the equal spin triplet pairs are only present in the magnetic domain region and can not spread to the other two ferromagnetic layers. It is hoped that our results could propose a new way to generate the long-range Josephson current.

   This work is supported by the State Key Program for Basic Research of China under Grants No. 2011CB922103 and No. 2010CB923400, and the National Natural Science Foundation of China under Grants No. 11174125 and No. 11074109.

    \end{document}